\input harvmac
\overfullrule=0pt
\def\Title#1#2{\rightline{#1}\ifx\answ\bigans\nopagenumbers\pageno0\vskip1in
\else\pageno1\vskip.8in\fi \centerline{\titlefont #2}\vskip .5in}

scaled\magstep3
 
scaled\magstep3

\overfullrule=0pt
\def\Title#1#2{\rightline{#1}\ifx\answ\bigans\nopagenumbers\pageno0\vskip1in
\else\pageno1\vskip.8in\fi \centerline{\titlefont #2}\vskip .5in}

scaled\magstep3
 
scaled\magstep3

\def\L{{\Lambda}}
\def\p{{\partial}}
\def\a{{\alpha}}

\def\e{{\epsilon}}
\def\apm{{\alpha^{\prime}}}
\lref\mr{R. Minasian and G. Moore, hep-th/9710230.}
\lref\cl{Y. E. Cheung and Z. Yin, hep-th/9710206.}
\lref\hmg{J. Harvey, G. Moore and M. Green, Class. Quant. Grav {\bf 14}
(1997) 47, hep-th/9605033.}
\lref\cvew{C. Vafa, Nucl. Phys. {\bf B447} (1995) 252.}
\lref\mms{K. Behrndt and  I. Gaida,
Phys. Lett. {\bf{B401}} (1997) 263, 
 hep-th/9702168;  J. Maldacena, G. Moore and A. Strominger, in progress.}
\lref\suss{L. Susskind and J. Uglum, Phys. Rev. {\bf D50} (1994) 2700.}
\lref\fpst{T. Fiola, J. Preskill, A. Strominger and S. Trivedi, 
Phys. Rev. {\bf D50} (1994) 3987.}
\lref\gkl{S. Gubser I. Klebanov, hep-th/9708005.}
\lref\cand{P. Candelas,P. Green, L. Parke and X. dela Ossa,
Phys. Lett. {\bf B258} (1991) 118.}
\lref\shm{M. Shmakova, Phys. Rev. {\bf D56} (1997) 540, hep-th/9612076.}
\lref\schwarz{J. Schwarz, hep-th/9601077.}
\lref\hms{G. Horowitz, J. Maldacena and A. Strominger,
Phys. Lett. {\bf B383} (1996) 151,
hep-th/9603109.}
\lref\jmtalk{ For a recent summary and references see J. Maldacena,
hep-th/9705078. } 
\lref\indx{Dim of mod space}
\lref\afmn{I. Antoniadis, S. Ferrara, R. Minasian and K.S. Narain,
hep-th/9707013.}
\lref\fks{S. Ferrara, R. Kallosh and A. Strominger, 
Phys. Rev. {\bf D52} (1995) 5412,
hep-th/9508072. }

\lref\cv{F. Larsen and F. Wilczek, Phys. Lett. {\bf B375} (1995) 37.}  

\lref\gs{M. B. Green and J. Schwarz, Nucl. Phys. {\bf B198} (1982) 441.}
\lref\grgu{M. Green and M. Gutperle, 
 Nucl. Phys. {\bf B498} (1997) 195,  hep-th/9701093; 
M. Green, M. Gutperle and P. Vanhove, 
hep-th/970617. }
\lref\wfv{E. Witten, 
hep-th/9610234.}
\lref\vff{M. Bershadsky, V. Sadov and C. Vafa,
Nucl. Phys. {\bf B463} (1996) 398,
hep-th/9510225.}

\lref\kleb{I. Klebanov and A. Tseytlin, 
Nucl. Phys. {\bf B475} (1996) 164,  hep-th/9604089.   }
\lref\jmas{J. Maldacena and A. Strominger, 
hep-th/9710014. }
\lref\jmfive{J. Maldacena, 
Nucl. Phys. {\bf B477} (1996) 168,
 hep-th/9605016. }
\lref\bm{K. Berndt and  T. Mohaupt,
  Phys. Rev. {\bf D56} (1997) 2206, hep-th/9611140.}
\lref\jmcy{J. Maldacena, 
Phys. Lett. {\bf B403} (1997) 20,
hep-th/9611163.} 
\lref\bcov{
M. Bershadsky, S. Ceccoti, H. Ooguri and C. Vafa, 
Nucl. Phys. {\bf B405} (1993) 279.
}
\lref\klms{
 D. Kaplan, D. Lowe, J. Maldacena and  A.
Strominger,
Phys. Rev.  D55  (1997) 4898,  hep-th/9609204.
}
\lref\cvewoneloop{
C. Vafa and E. Witten, {\it A one loop test of string 
duality}, Nucl. Phys. {\bf B447} (1995) 261,  hep-th/9505053. ???????  }
\lref\vf{C.Vafa,}
\lref\kb{K. Behrndt, G. Lopez Cardoso, B. de Wit, R. Kallosh, 
D. Lust and T. Mohaupt, 
Nucl. Phys. {\bf B488} (1997) 236,  hep-th/9610105. }
\lref\asm{A. Strominger, 
 Phys. Lett. {\bf B383} (1996) 39,  hep-th/9602111.  }
\lref\ascv{A. Strominger and C. Vafa, 
 Phys. Lett. {\bf B379} (1996)
99,
 hep-th/9601029.}
\lref\berk{N. Berkovitz, 
hep-th/9709116. }

\lref\hlm{ G. T. Horowitz, D. A. Lowe and J. M. Maldacena,
Phys. Rev. Lett. 77 (1996) 430, hep-th/9603195.}
 
\Title{\vbox{\baselineskip12pt
\hbox{hep-th/9711053}}}
{\vbox{\centerline{\bf{ BLACK HOLE ENTROPY IN}}
\vskip2pt\centerline{\bf{M-THEORY}}}}

\baselineskip=12pt
\centerline{ Juan Maldacena, Andrew Strominger }
\smallskip
\centerline{\sl Department of Physics}
\centerline{\sl Harvard University}
\centerline{\sl Cambridge, MA}
\smallskip
\centerline{and}
\smallskip
\centerline{Edward Witten}
\smallskip
\centerline{\sl The Institute for Advanced Study}
\centerline{\sl Princeton, NJ 08540}

\bigskip
\centerline{\bf Abstract}
{ Extremal black holes in M-theory compactification on 
$M\times S^1$ are microscopically
represented by fivebranes wrapping $P\times S^1$, where $M$ is a 
Calabi-Yau threefold and $P$ is 
a four-cycle in $M$. Additional spacetime charges arise from momentum 
around the $S^1$ and expectation values for the self-dual three-form 
field strength in the fivebrane. The microscopic entropy of the fivebrane 
as a function of all the charges is determined from a two-dimensional 
$(0,4)$ sigma model whose target space includes the fivebrane moduli space. 
This entropy is compared to the macroscopic formula. Precise agreement 
is found for both the tree-level and one-loop expressions.}

\Date{}
\newsec{Introduction}
Thanks to recent advances in string theory, the thermodynamic properties 
of black holes can be microscopically 
derived in a variety of contexts with remarkable precision \jmtalk. This has 
led to new insights into the quantum structure of black holes 
as well as  string theory itself. The macroscopic-microscopic
correspondence has developed to the extent that semiclassical black holes
have become a powerful tool for analyzing the quantum theories 
which describe their dynamics \refs{\kleb \jmfive \gkl -\jmas}.  

In this paper 
we will compute and compare the 
macroscopic and microscopic entropies of extremal black holes arising 
in an M-theory compactification to four dimensions on $M\times S^1$,
where $M$ is a Calabi-Yau threefold, 
so that the unbroken spacetime supersymmetry is precisely
$N=2$.  Some special cases were considered in \klms\ and a 
heuristic explanation for the  more general case was attempted in 
\refs{\bm - \jmcy}.

Several new features arise in the 
computation. The basic microscopic object which enters is the 
still-mysterious M-theory fivebrane, aspects of 
whose world-volume field theory enter the analysis. 
The resulting geometric picture is quite interesting, since
it maps the black hole degrees of freedom to the different ways of 
deforming a ``foam'' of fivebranes.
  We are able to go beyond 
the tree-level Bekenstein-Hawking 
entropy and successfully compute and compare the one-loop corrections. The 
macroscopic correction at one loop is related to the entropy arising from
entanglement 
of the quantum state inside and outside the horizon,
 while at the microscopic level it is a 
subleading contribution to the central charge of a two-dimensional
conformal
 field theory.
A curious feature of the analysis is that we never need to  
use string theory itself: everything follows from the properties of 
M-theory fivebranes in curved spaces. Indeed with 20/20 hindsight, 
this paper might have 
been written years ago following the discovery of eleven-dimensional 
supergravity, Calabi-Yau compactification, and the fivebrane as a spacetime
soliton.

This paper is organized as follows. In section 3.1 we review the 
macroscopic entropy formula, and the explicit 
solution for general nonzero $S^1$ momentum, twobrane and fivebrane 
charges. In section 3.2 we compute the leading 
one loop correction which arises from the $R^4$ 
term in eleven-dimensional M-theory.  In section 3 we describe the 
$(0,4)$ sigma model that counts the BPS states. In 3.1 we compute the
central charge in terms of the homology class of the four cycle $P$. 
In 3.2 we relate antisymmetric tensor expectation values to twobrane
charges and show that this affects the entropy via a shift in the effective
$S^1$ momentum. In section 4 we briefly raise - but do not resolve - the
issue of $\apm$ corrections in the string theory regime.

\newsec{Macroscopic Entropy}
\subsec{The Area Formula}  
In this subsection, we review the semiclassical area-entropy formula 
for an $N=2,~d=4$ extremal black hole characterized by 
magnetic and electric charges $(p^\L,q_\L)$. The asymptotic values of the 
vector moduli $Z^\L=X^\L/X^0, ~\L=0,1,\dots,n_V$, in the black 
hole solution are arbitrary. These moduli couple
to 
the electromagnetic fields and accordingly vary
as a function of the radius. At the horizon they approach a fixed point whose
location in the moduli space depends only on the charges \refs{\fks,\cv}. The 
locations
of these fixed
points
can be found by looking for supersymmetric 
solutions with constant moduli. The general
formula is \refs{\asm,\fks} \foot{ We have redefined $C, ~X$ and $F$ by factors
of $e^{K/2}$ relative to \asm\ so that $(X^\L,F_\L)$ is a
holomorphic section. } 
\eqn\prf{p^\L=Re[CX^\L],}
\eqn\qrf{q_\L=Re[CF_\L],}
where $F_\L=\p_L F$ are the holomorphic periods. The $2n_v+2$ real 
equations \prf\ and  \qrf\ determine the $n_v+2$ complex quantities $(C,X^\L)$ 
up to Kahler transformations. The Kahler potential $K$
is given by 
\eqn\cgg{X^\L\bar F_\L-\bar X^\L F_\L=ie^{-K}.}
Given the horizon values of the
moduli determined by \prf\ and
\qrf\ the Bekenstein-Hawking entropy is 
\eqn\ent{S_{BH}={1 \over 4}{\rm Area}
={\pi iC\bar C \over 4}(X^\L \bar F_\L-\bar X^\L F_\L).}

For M-theory compactified on $M\times S^1$, where $M$ is a Calabi-Yau 
threefold, the prepotential is 
\eqn\fcy{F(X)=D_{ABC}{X^AX^BX^C \over X^0}}
where $A,B=1,..n_V$. The intersection numbers $6D_{ABC}$ are
\eqn\dabc{6D_{ABC}\equiv \int_M \a_A\wedge\a_B\wedge\a_C ,}
where the $\a_A$ are an integral basis for $H^2(M;{\bf Z})$. 

The fixed point equations \prf\ \qrf\ have been solved for 
a general prepotential of the form \fcy\ 
with the restriction $p_0=0$ \refs{\shm, \kb}. The fixed 
points are given by
\eqn\fpt{\eqalign{CX^0&={i}\sqrt{D \over \hat q_0},\cr
     CX^A&=p^A+
{i\over 6}\sqrt{D \over \hat q_0}D^{AB}q_B,\cr}}
where 
\eqn\ddf{\eqalign{D & \equiv D_{ABC}p^Ap^Bp^C, \cr  
\hat q_0 & \equiv q_0+{1 \over 12}D^{AB}q_Aq_B, \cr
     D_{AB} & \equiv D_{ABC}p^C, \cr
      D^{AB}D_{BC}&=\delta^A_{~~C}. \cr}}
The entropy then follows from \ent\ and \fcy\ as 
\eqn\qsf{S=2\pi\sqrt{D\hat q_0}.}

In order that the long wavelength approximation to
M-theory can be trusted, the volume  $V_M$ of 
$M$ as well as the radius $R$
of the $S^1$ should both be large in eleven-dimensional Planck units. 
$V_M$ is a hypermultiplet scalar which is 
a freely adjustable constant throughout the black hole solution. 
$R^3V_M$ is a scalar in a vector multiplet whose value at the horizon 
is proportional to 
$\sqrt{\hat q_0^3/D}$. Hence the validity of the semiclassical  M-theory 
computation requires $\hat q_0^3 \gg D$.
The validity of the long wavelength approximation
 requires not just that the total volume of $M$ should be large
 at the horizon, but that the volume of any two-cycle in $M$ should
 be large there.
 It follows from \fpt\ that the  Kahler class at the horizon
  is  proportional to 
$P\equiv p^A\a_A$. Hence we require that $P$ lies inside the Kahler cone,
a restriction that will be extensively used later.

The above conditions were necessary to ensure that we can derive 
the effective four dimensional theory by simple Kaluza-Klein reduction. 
Demanding that the black hole  supergravity solution is weakly curved
further requires
\eqn\sugracond{ D \gg V_M. }

\subsec{Macroscopic Loop Corrections}

The macroscopic entropy \qsf\ was derived from the classical
leading low-energy effective $d=4, ~N=2$ supergravity action.
This action has corrections from higher-dimension operators.
Corrections to the underlying eleven-dimensional action 
are a power series in the Planck length while 
corrections to the four dimensional action also involve 
$R$ and $V_M$. 
In general there is no known systematic procedure for 
computing these corrections in the M-theory regime of large $R$ and 
large $V_M$. However, the correction we are after arises from 
a special term whose coefficient can be determined.
This is the $R^2$ correction 
\eqn\coril{S_1={1 \over 96 \pi}\int c_{2A} {\rm Im}Z^A R\wedge *R ,}
where
\eqn\zx{Z^A(x)={X^A \over X^0},}
are moduli fields and
\eqn\cdf{c_{2A}\equiv \int_M c_2(TM)\wedge \alpha_A ,}
with $c_2$ the second Chern class of $M$. 
This term has a topological origin \refs{\bcov \cvew -\afmn} 
and descends from the $R^4$ term in $d=11$ \afmn. In string
perturbation theory -- that is, in an expansion in $1/V_M$ -- this $R^4$ term 
has been computed as a one loop correction \gs. It cannot 
be renormalized
at higher loops because it is related to 
anomaly cancellation\refs{\grgu, \berk}.

In order to determine the effect of \coril\ 
on the entropy we first consider the special 
case that the moduli fields take their 
constant fixed point values throughout 
the black hole solution. It then 
follows from
\fpt\ that ${\rm Im}Z^A=-p^A \sqrt{\hat q^0/D}$. Using the fact that the 
Euclidean black hole solution is $R^2\times S^2$ and has Euler character
2, \coril\ is then 
\eqn\sdb{S_1=-c_{2A}  p^A {\pi \over 6}\sqrt{\hat q^0 \over D}.}
This correction to the effective action is a correction to $-\log Z = \beta
F $, so the correction to the entropy will be
$ \delta S = \delta \log Z - \beta \partial_\beta \delta \log Z $.
Since the correction \coril\ is topological (proportional to 
the Euler character), it will be independent of the temperature.
The correction to the entropy is then
\eqn\bbn{\Delta S=-S_1= c_{2A}  p^A{\pi \over 6}\sqrt{\hat q^0\over D}.}
Note that unlike the leading term \qsf, this term is of order zero, rather
than quadratic in the charges. It will, however, be large in the case 
we are considering where  $q^0 \gg D^{1/3} $.

In general, the value of $Z^A$ at infinity is arbitrary, and $Z^A$ will vary
over the black hole solution. The microscopic entropy
is a function only of the charges and is independent of the asymptotic values
of the the moduli, because the number of BPS states cannot vary smoothly
with the moduli. Hence the macroscopic entropy should also 
be independent of the moduli. The macroscopic origin of this in the spacetime
solutions has been understood for 
tree level entropy in terms of fixed points as described above. 
We do not understand the macroscopic mechanism 
for moduli-independence of the one loop corrections 
for the general case when the moduli fields are not taken to be at 
their fixed points. This will require a detailed knowledge 
of the one loop corrections and supersymmetry transformation
laws.

The one-loop correction to the effective action of course 
involves more terms than just \coril. Some of these involve gauge fields 
which are nonzero for the black hole solution, and in principle 
might further correct the entropy. However the topological nature 
of \coril\ suggests that it should play a dominant role,
and we shall indeed find that the induced correction to the entropy 
agrees with the microscopic prediction\foot{There are also subleading 
logarithmic corrections \mms\ not considered here.}. 

\newsec{Microscopic Entropy}

\def\Z{{{\bf Z}}}

At the microscopic level,  configurations with charges $(0,p^A,q_0,q_A)$
are obtained from a fivebrane wrapping the five-cycle $P\times S^1$
 and carrying 
total momentum $q_0$ about the $S^1$.
Here $P$ is a four-cycle
in $M$.  
Its cohomology 
class $[P]\in H^2(M,\Z)$  
can be expanded as $[P]=\sum_Ap^A\Sigma_A$ where
the  $p^A$ are charges and $\Sigma_A$ are a basis of $H^2(M,\Z)$;
one can consider the $\Sigma_A$ to be the  cohomology classes of a basis
of four-cycles in $M$.
Nonzero $q_A$ charge arises, as we discuss later,
from exciting the self-dual antisymmetric tensor 
field on the fivebrane. Nonzero $p^0$ -  not considered here - would arise 
from 
a Kaluza-Klein monopole on the $S^1$.

Supersymmetry requires that $P$ should be holomorphic; it is
of complex codimension one in $M$.  We will need to work out the low
energy effective field theory obtained by wrapping a fivebrane on
$P\times S^1$.  As explained at the end of section 2.1, $[P]$ is 
proportional to 
the effective Kahler class of $M$ near the black hole horizon, and hence
long-wavelength M-theory is a good description only if $[P]$ is inside
the Kahler cone in $H^2(M,\Z)$, so that $M$ is smooth at the horizon,
 and moreover is large.
These conditions mean that $P$ is a ``very ample divisor'' in the language
of algebraic geometry, 
and lead to many simplifications.  One important simplification
is that we can assume that $P$ is smooth; in the moduli space of $P$'s,
one will encounter singularities at special points, but the generic
$P$ of given cohomology class is smooth.  The fact that $P$ is generically
smooth
means that we do not need to understand the behavior of M-theory
with coincident or intersecting fivebranes; everywhere there is locally
a single, isolated, smooth fivebrane.

Another important consequence of the fact that $P$ is very ample
is that there is a general method to describe the moduli of $P$.
Given any complex divisor $P$ in a complex manifold $M$, there is
a holomorphic line bundle ${\cal L}$ and a holomorphic section $s$ of
${\cal L}$ such that $s$ vanishes precisely along $P$.   Existence
of such an $s$ means that $c_1({\cal L})=[P]$.  Conversely,
any holomorphic section $s'$ of ${\cal L}$ vanishes along a divisor $P'$
whose cohomology class equals that of $P$.  $s'$ is uniquely determined
by $P'$ up to scaling by a complex constant ($s'\to\lambda s'$ with
$\lambda\in {\bf C}^*$).  So the moduli space ${\cal M}$ of divisors
that are cohomologous to $P$ is a complex projective space that
is obtained by projectivizing the vector space $H^0(M,{\cal L})$.

This statement holds for any divisor $P$ in a complex manifold.
For a general $P$, there would be no nice formula for the dimension
of $H^0(M,{\cal L})$.  There is, however, a nice index or Riemann-Roch
formula for the alternating sum 
$w=\sum_{i=0}^{{\rm dim}\,M}(-1)^i
{\rm dim}\,H^i(M,
{\cal L})$.  In fact,
\eqn\hubb{w=\int_M e^P\,{\rm Td}(M).}
Here ${\rm Td}(M)$ is the Todd class of $M$.  Also, we
have used the fact that $c_1({\cal L})=[P]$,
and we henceforth sometimes write simply $P$ instead of $[P]$ for
ease of reading the formulas. 
For a divisor in a Calabi-Yau threefold, one has
${\rm Td}(M)=1+c_2(M)/12$, with $c_2(M)$ the second
Chern class.  So the index formula can be evaluated to give
\eqn\bubble{w=\int_M\left(
{P^3\over 6}+{1\over 12}Pc_2(M)\right).}

In \bubble, there is no requirement of ampleness.
But for  $P$ very ample, ${\rm dim}\,H^i(M,
{\cal L})=0$ when $i > 0$.   So in this case, the definition 
of $w$ reduces to $w={\rm dim}\,H^0(M,{\cal L})$.
The quantity $w$ in \bubble\ is accordingly for very ample $P$
the number of complex parameters required
to determine a holomorphic section $s$ of ${\cal L}$.  Because of
the equivalence under $s\to \lambda s$, the number of complex moduli
of $P$ is $w-1$, so the number of real moduli of $P$ is $2w-2$ or
\eqn\dsp{d_p=\int_M({1 \over 3}
P^3+{1 \over 6}Pc_2(M))-2.}

These moduli can vary as we move along the $S^1$. We take the radius $R$
of the $S^1$ much bigger than the typical size of the Calabi Yau space, in the
sense that
$ R^6 \gg V_M$.  The low energy dynamics is then described by 
a two-dimensional sigma model on the $S^1$ as in \ascv. This sigma model 
has $(0,4)$ supersymmetry. The chirality of the sigma model descends
from the $(0,2)$ chirality of the field theory on the fivebrane.  
This sigma model contains the moduli of the four-cycle, as well 
as scalar fields corresponding to expectation values of the 
two index antisymmetric tensor potential that lives on the fivebrane 
world volume. 
As in \ascv, the microscopic entropy is given by  the logarithm of 
the number of left-moving excitations with total momentum $q_0$.
For large $q_0$ ($q_0\gg c_L$), this asymptotically approaches
\eqn\icrent{S_{micro}=2\pi\sqrt{c_L\tilde q_0 \over 6},}
where $c_L$ is the left-moving (non-supersymmetric) central charge 
and $\tilde q_0$ is the momentum available to freely distribute 
among the left-moving oscillators. 

The above discussion makes sense when the fivebranes are far apart from
each
other so that we can neglect their gravitational effects, and also when
they 
are embedded in a large space where we keep just the low energy description
of the fivebrane. This requires that $ D \ll V_M$, which is just the 
opposite of the condition \sugracond\ for the validity of the 
supergravity analysis.  Since $V_M$ is in a
hyper multiplet, we can presumably interpolate between these two regimes 
without changing the number of BPS states. 
In order to reduce the dynamics to the
sigma model we needed that $R^6 \gg V_M$, which will require changing 
the value of a vector multiplet. We are assuming throughout 
that there are
no jumping phenomena (or that they can be neglected) when we perform this
change.

\subsec{Computing $c_L$}
In this section we compute the microscopic entropy for the case 
of large $q_0$, so that we may approximate $q_0=\tilde q_0=\hat q_0$. 
The shift in $q_0$ will be considered in the next section. 
The results will reproduce for this case the 
leading entropy \qsf\ as well as the one-loop semiclassical correction 
\bbn.  

One contribution to $c_L$ comes from massless fields that arise
from fluctuations in the moduli of $P$.  Such fields propagate
as left- and right-movers on the $S^1$, and the left-movers contribute
to $c_L$.  There are no fermions contributing to $c_L$ for the following
reason.  Fermions arise from  
$(0,i)$ forms on $P$. They are right-moving
for $i$ even and left-moving for $i$ odd. Hence the number of left-moving
fermions is $b_1(P)$, the first Betti number of $P$.  For $P$ a very
ample divisor in a Kahler manifold $M$, the Lefschetz hyperplane
theorem says that $b_1(P)=b_1(M)$.  For $M$ a complex threefold whose holonomy
is $SU(3)$ (and not a subgroup), $b_1(M)=0$, and so there are no
left-moving fermions.\foot{If the holonomy of 
$M$ is a proper subgroup of $SU(3)$ -- in other
words, if $M$ is a six-torus or a two-torus times K3 -- then $b_1(M)\not=0$,
and there would be left moving fermions. For such $M$, the
unbroken supersymmetry is greater than the $N=2$ assumed in the present
paper, and a few other details of the exposition would also be modified.}

The other bosonic contribution to $c_L$ comes from the 
 the rank two antisymmetric
tensor potential $b$ that propagates on the fivebrane world-volume.
To determine its dimensional reduction to $S^1$, it is important
that the three-form field strength $h=db$ is self-dual.
The reduction of $b$ to $S^1$ gives $b_2^+$ right-moving
massless scalars on $S^1$ and $b_2^-$ left-moving ones, where $b_2^+$
and $b_2^-$ are the dimensions of the space of self-dual and
anti-self-dual two-forms on $P$.  
 (The reduction of $b$ does
not give two-dimensional gauge fields, since the first Betti number
$b_1(P)$ vanishes for a reason given in the next footnote.)

The Euler characteristic $\chi$ and signature $\sigma$ of $P$
can be expressed in terms of $b_2^{\pm }$ by $\sigma=b_2^+-b_2^-$ and
$\chi=2+b_2^++b_2^-$.\foot{
In general, for a four-manifold $P$, $\chi(P)=2-2b_1(P)+b_2^++b_2^-$,
where $b_1(P)$ is the first Betti number of $P$, but we assume here that 
$M$ is a complex threefold whose holonomy
is $SU(3)$ (and not a subgroup). In this case, as discussed above,
$b_1(M)=b_1(P)=0$, 
so the formula for the
Euler characteristic of $P$ reduces to $\chi=2+b_2^++b_2^-$.}
$\chi$ and $\sigma$ can be computed as follows.
It suffices to know the Chern classes $c_1(P)$ and $c_2(P)$, since
for a two-dimensional complex manifold $P$ one has
\eqn\eeqer{\eqalign{\chi & =\int_Pc_2(P),  \cr
                    \sigma & = -{2\over 3}\chi+{1\over 3}\int_Pc_1(P)^2.\cr}}
                    
The Chern classes of $P$ can be computed in terms of the Chern classes
of $M$ and the cohomology class $[P]$.  Let $TP$ and $TM$ be the tangent
bundles to $P$ and $M$, so that by definition $c_i(P)=c_i(TP)$,
$c_i(M)=c_i(TM)$.  Let $TM|_P$, ${\cal L}|_P$ denote
the restrictions of $TM$ and ${\cal L}$ to $P$.  There is an
exact sequence
\eqn\orbo{0\to TP\to TM|_P\to {\cal L}|_P\to 0,}
which expresses the fact that the restriction of ${\cal L}$ to $P$
can be understood
as the normal bundle to $P$ in $M$.\foot{For instance, if $s$
is the section of ${\cal L}$ that vanishes along $P$, and $s'$ is any
other section, then the vanishing of $s+\epsilon s'$ defines
a divisor $P_\epsilon$
that is homologous to $P$. To first order in $\epsilon$, the 
$\epsilon$-dependence of $P_\epsilon$ describes a first order displacement
of $P$ which can be understood as a section of its normal bundle;
so the section $s'$ of ${\cal L}$, restricted to $P$,
 can be interpreted as a section of
the normal bundle to $P$ in $M$.}
Hence, if $c=1+c_1+c_2+\dots$ is the total Chern class, we have
\eqn\fluffo{
c(TM|_P)=c(TP)c({\cal L}|_P).}  Because $M$ is Calabi-Yau,
we have $c_1(TM)=0$, and hence $c(TM|_P)=1+c_2(M)+\dots$.
So \fluffo\ gives the relations $c_1(P)=-c_1({\cal L})=-[P]$, and
$c_2(P)=c_2(M)+[P]^2$.\foot{Since $c_i(P)$ is defined as a cohomology class of
$P$, all classes appearing on the right hand side of these formulas should
be restricted to $P$.  We do not indicate this in the notation, and in
any event will momentarily extend the classes away from $P$. 
}
With these relations, we can use \eeqer\ to express
$\chi$ and $\sigma$ in terms of $\int_P c_2(M)$ and $\int_P[P]^2$.

It is now  convenient  to note that if $A$ is any cohomology
class on $M$, whose restriction to $P$ we write also as $A$, then
$\int_PA=\int_MP\cdot A$.  Using this principle,
we can write $\chi(P)$ and $\sigma(P)$ in terms of integrals on $M$.
Writing simply $P$ instead of $[P]$, we get
\eqn\cchh{\eqalign{\chi&=\int_P c_2(P)\cr
                       &=\int_P (c_2(M)+c_1^2(P))\cr
                        &=\int_M(Pc_2(M)+P^3),}}
\eqn\signa{\eqalign{\sigma&=-{2 \over 3}\chi +{1 \over 3}\int_P
 c_1(P)^2 \cr
                         &=-\int_M({1 \over 3}P^3+{2 \over 3}Pc_2(TM)).}}

Expressing $b_2^\pm$ in terms of $\chi$ and $\sigma$
by $b_2^\pm=\half(\chi \pm \sigma)-1$ 
yields
\eqn\btm{b_2^-=\int_M\left({2 \over 3}P^3+{5 \over 6}Pc_2(TM)\right)-1,}
\eqn\btp{b_2^+=\int_M({1 \over 3}P^3+{1 \over 6}Pc_2(TM))-1.}

The total number of left- and right-moving massless
bosons is $N^B_L=d_p+b_2^-+3$, $N^B_R=d_p+b_2^++3$, where in each case $+3$
is the contribution of three translational zero modes.
So in terms of 
\eqn\cpd{c_2 \cdot P = \int_M Pc_2(TM), }
and 
\eqn\drv{D={1 \over 6}\int_M  P^3,}
we get 
\eqn\clt{c_L=N_L^B = 6D+c_2 \cdot P,}
\eqn\clr{c_R=N_R^B + {1 \over 2 } N_R^F = 6D+{1 \over 2}c_2 \cdot P.}
Here we have included in $c_R$ also $2 D + c_2\cdot P/6 $ 
 complex  fermions built out
of $(0,2)$ forms as required by supersymmetry (there are no left moving
fermions
since $b_1=0$).
Note that in comparing to section 2, $D$ should be identified with the
quantity of the same name introduced in equation \ddf.  The $p^A$ in \ddf\
are the expansion coefficients of $[P]$ in a basis of classes $\Sigma_A$;
thus $[P]=\sum_Ap^A\Sigma_A$.  In the basis $\Sigma_A$, the intersection
form of $M$ is given by $6D_{ABC}=\Sigma_A\cap \Sigma_B\cap \Sigma_C$.

The microscopic entropy  for large $q_0$
is then from \icrent
\eqn\smic{S_{micro}=2\pi \sqrt{(6D+c_2 \cdot P)q_0\over 6}.} 
To compare with \qsf\ we should expand in powers of $1/D$. 
One finds
\eqn\cnp{S_{micro}=2\pi\sqrt{Dq_0}+c_2\cdot P {\pi \over 6}\sqrt{q_0 \over
D}+ . . .}
This agrees with the Bekenstein-Hawking result \qsf\ plus the one loop 
correction \bbn. It would be interesting to macroscopically reproduce 
the full series of corrections obtained microscopically in \smic. 

The computation given here for Calabi-Yau spaces
can be repeated for ${\rm K3} \times T^2 $
or $T^6$ with only minor differences. 
Again the generic fivebrane configuration is described by a smooth
holomorphic
map. At special points in the moduli space it degenerates into 
a set of fivebranes with $D$ intersections. We can view the moduli 
as coming from ``blow-up'' modes associated to each
intersection. In all cases we find that 
the sigma model is (0,4) if $D$ is nonzero. 
For the  $K3 \times T^2 $
or $T^6$ case there are additional 
modes coming from $b_1$, which are subleading for large $D$.

\subsec{ Membrane Charge} 

In this section we consider the effect of endowing our black
hole with  nonzero membrane charge $q_A$. 
Membrane charge is actually carried in the effective two-dimensional
theory by the massless scalars that arise from dimensional reduction
of the chiral two-form $b$; they were counted in section 3.1.
The reason for this is that membrane charge is the flux of the self-dual
three-form $h=db$ over cycles of the form $S^1\times \Sigma$,
with $\Sigma$ a two-cycle in $P$.  The fluxes for various $\Sigma$
reduce in terms of the low energy physics on $R\times S^1$ to the winding
numbers around $S^1$ of the various massless scalars  that arise from
reduction of $b$. It can be seen that all membrane charges arise in this 
fashion if $P$ is  in the Kahler cone.  

Thus, the membrane charge is a vector in the Narain lattice of the
massless scalars.  Now, actually there are two lattices of interest.
The obvious lattice is $\Gamma=H^2(P,{\bf Z})$.  It has a sublattice
$\Gamma_M=H^2(M,{\bf Z})$, consisting of two-dimensional classes on $P$
that can be extended over $M$.\foot{$\Gamma_M$ is a sublattice of
$\Gamma$ because for $P$ a very ample divisor in $M$, the restriction
map from $H^2(M,{\bf Z})$ to $H^2(P,{\bf Z})$ is injective.  This is
so because an ample divisor $P$ has a positive intersection with every
divisor $D$ in $M$; in fact, the triple intersection number  $D \cap P
\cap P$ is positive, being the volume of $D$ in a Kahler metric whose
Kahler class is $P$.} 
The lattice $\Gamma$ has signature $(b_2^+,b_2^-)$, with $b_2^\pm$
computed in section 3.1.  The lattice $\Gamma_M$ has signature
$(1,b_2(M)-1)$.  The last statement is proved as follows.  Since
$M$ has holonomy $SU(3)$ (and not a proper subgroup of $SU(3)$),
$H^{2,0}(M)=0$.  Hence $H^2(M,{\bf R})$ is generated by differential
forms of type (1,1), and the sublattice $\Gamma_M$ is entirely of type (1,1).
By the Hodge index theorem, $H^{1,1}(P)$ has a self-dual part that is
one-dimensional, generated by $[P]$ itself, so the positive signature
part of $\Gamma_M$ is at most one-dimensional.  Since $[P]$ does extend over
$M$, it is a vector in $\Gamma_M$, and hence $\Gamma_M$ has signature
$(1,b_2(M)-1)$ with the positive signature subspace being generated by
$[P]$.

At first sight it seems that, for a fivebrane wrapped on $S^1\times P$,
the membrane charge should be a vector in $\Gamma$.  Actually,
we should take it to be a vector in the sublattice $\Gamma_M$.  There
are two closely related reasons for this.

(1) From the point of view of the underlying M-theory on
$R^4\times S^1\times M$, the conserved membrane charges are vectors
in $\Gamma_M$, not in the larger lattice $\Gamma$.  Thus, in discussing
the macroscopic entropy in section 2, the membrane charge was a vector
in $\Gamma_M$.  We should expect that any fivebrane state, with a charge
vector that initially is in $\Gamma$ but not in $\Gamma_M$, can decay to 
a state with charge vector in $\Gamma_M$.  (In this decay, the inner
product of the charge vector with any vector in $\Gamma_M$ will be conserved.)

(2) To build a BPS state, the membrane charge vector must be in $\Gamma_M$.
In fact, the membrane charge vector must be integral, and 
the BPS condition requires that the membrane charge should be
a sum of a left-moving vector and a multiple of $P$.
This important statement will be justified at the end of the present
section.
Because $\Gamma_M$ has signature $(1,b_2(M)-1)$ with the plus part
generated by $[P]$, a charge vector in $\Gamma_M$ automatically
obeys the necessary conditions for a BPS state.  Generically, a vector
not in $\Gamma_M$ would not obey those conditions.  The reason for the last
assertion is that for $P$ a very ample divisor in $M$, one has $H^{2,0}(P)
\not= 0$,\foot{In fact, by the Hodge index theorem, $b_2^+=1+2\,{\rm dim}\,
H^{2,0}(P)$.  The formula for $b_2^+$ in section 3.1 thus  gives
a formula for ${\rm dim}\, H^{2,0}(P)$.}
which generically gives an obstruction to the existence of
vectors not in $\Gamma_M$ and obeying the desired conditions for
a BPS state.

In the macroscopic discussion of section 2, 
the effect of such nonzero membrane charges on the macroscopic 
entropy is simply to shift
$q_0$ in equation \qsf.  From a microscopic point of view,
this happens because membrane charge -- in other words, a nonzero winding
number of the scalars -- shifts the ground state energy and momentum
of the effective two-dimensional theory.  This can be seen as follows.

We recall that 
the conformal field theory on $R\times S^1$ has one chiral 
boson, arising from reduction of 
the self-dual antisymmetric tensor $b$ on the fivebrane, for every harmonic 
two-form on $P$. 
Since we wish to consider the case of a membrane charge vector in $\Gamma_M$, 
we focus on contributions from  
 two-forms $\alpha_A$ on $P$ which arise as 
the restriction to $P $ of closed 
two-forms $\alpha_A$ on $M$ (we denote them with the same letter).
  For simplicity we assume that the
Kahler class $J$ of $M$ is a multiple of $P$.
 This need not be the case in general, but 
the index of BPS states  does not depend on the choice of $J$, so it suffices
to consider this case.  In any event, we have seen in section 2 that
at the black hole horizon, $J$ is a multiple of $P$.

We obtain  two-dimensional fields $\phi^A$ by an ansatz $b=\sum_A
\phi^A \alpha_A$ for the two-form potential.
The  fields $\phi^A$ are  constrained by self-duality 
of the three-form field strength $h$ in $R\times S^1\times P$. In fact,
\eqn\phc{P^A_{\mp B}\partial_\pm \phi^B=0,}
where the projection operators $P_\pm$ are
\eqn\ppm{P^A_{\pm B}=\half (\delta^A_{~~B}\pm {1 \over 6}D^{AC}g_{CB}),}
with 
\eqn\bsb{g_{AB}=\int_P\alpha_A \wedge *\alpha_B=- 6 D_{AB}+
{12 D_{AC}p^CD_{BE}p^E\over D},} 
and $*$ is the Hodge dual in $P$. Let $k^A$ be the winding numbers of
the $\phi^A$, and let $k^A_\pm=P^A_{\pm B}k^B$. 
For nonzero $k$ there is a zero-mode 
contribution to the $S^1$ momentum given by
\eqn\zmm{{\Delta q_0 \over R} = \int_{S^1} d\sigma (T_{--}-T_{++})=  2 \pi^2 R 
g_{AB}(k_+^Ak_+^B-k_-^Ak_-^B)= - 12 \pi^2 R  D_{AB}k^Ak^B .}

We will now demonstrate more precisely that
 a state with nonzero momentum $k^A$ has 
a nonzero flux of $h$ and therefore caries membrane
charge.   The three-form 
potential $A^{(3)}$ of the low energy eleven-dimensional supergravity
has couplings both to the world volume of a membrane, which we will
take to be $R\times Q$, where $R$ is parametrized by time and $Q$ is
a two-surface in space, and to
the five-brane worldvolume $R\times S^1\times P$.
The couplings are
\eqn\asr{
\int_{R\times Q}A^{(3)}+ \int_{R\times S^1\times P} h \wedge A^{(3)}.}
The three form gauge potential gives rise to 3+1-dimensional 
$U(1)$ gauge fields
$A^A_\mu ,
~\mu=0,..3$ via the 
decomposition
\eqn\decomp{
A^{(3)} = A^A_\mu dx^\mu \wedge \alpha_A.
}
Membrane charge acts as a source for these gauge fields. 
When $k$ is nonzero, $h$ becomes
\eqn\hed{h=\alpha_Ak^A_{+}\wedge dx^+ +\alpha_Ak^A_{-}\wedge dx^-.}
It follows from \asr\ 
and the relation
\eqn\pws{\int_P\alpha_A\wedge \alpha_B= 6 D_{AB}}
that the effect of nonzero $h$ is to induce a membrane charge 
\eqn\mch{q_A=12\pi R  D_{AB}(k^B_+ - k^B_-).} 
Hence one finds that the total {\it leftmoving } 
$S^1$  momentum is  
\eqn\mes{q_0=\hat q_0-{1 \over 12}D^{AB}q_Aq_B,}
where $\hat q_0$ is the momentum carried by non-zero modes of 
the $c_L \approx 6 D$ conformal field theory and we have 
used $k^A_+k^B_- D_{AB} =0$. 
Since the entropy counts the number of ways of distributing the momentum 
within these modes, the microscopic entropy is 
\eqn\smc{S=2\pi\sqrt{D\hat q_0},}
in full agreement with \qsf , including the coefficient and the sign.

We are interested in supersymmetric BPS states and so we still 
need to check 
that the momenta do not break supersymmetry. All BPS states have the 
same number 
of supersymmetries as the $k=0$ ground state. Supersymmetry is 
realized by the right movers, so naively supersymmetry is broken 
whenever the right-moving momentum
 is nonzero. However, there is a crucial subtlety here.
In fact, if  the right-moving momentum
 is nonzero but is a multiple of $[P]$, then  all the 
linearly realized supersymmetries of the vacuum are broken but an equal 
number of unbroken supersymmetries 
reappear as combinations of the original linear and nonlinear 
supersymmetries, as follows.  The two-form $b$ on the membrane
world-volume has one distinguished right-moving mode coming from
$b\sim [P]$.  Let $k_R$ be the momentum of this mode.
This mode is
 paired together with the three translational zero modes and the 
four goldstinos $\psi$ in a $(0,4)$ supermultiplet. 
For nonzero $k_R$ (but no spacetime momentum), the $\psi$ 
transform under the linear supersymmetries as 
\eqn\pstr{\delta\psi=k_R\gamma^4\e^{lin},}
signalling the breaking of the original supersymmetries.
However as goldstinos 
they also transform under the nonlinear supersymmetries as
\eqn\gst{\delta\psi=\e^{nlin}.}
Hence linear combinations of transformations obeying
\eqn\scc{\e^{nlin}=-k_R\gamma^4\e^{lin}}
provide four unbroken supersymmetries for generic values of $k_R$, and 
BPS states exist for all $k_R$.  Such mixing of linear and nonlinear
supersymmetries in describing BPS states is of course familiar in many
other aspects of brane physics, for instance in matrix theory.

The other right-moving scalars of the theory are not
paired with the goldstinos in this way.  So their momenta must vanish
in a BPS state.  Hence, the BPS condition asserts
not that the right-moving charge is zero but that it is a multiple of
$[P]$, as we asserted near the beginning of section 3.2
in explaining why the charge vector
should lie in $\Gamma_M$.


\newsec{$\apm$ corrections}

In the preceding, we have considered corrections to the entropy 
which correspond to string loops in the type IIA picture.
These come from  higher dimension operators, and the leading correction 
is suppressed by a factor of $D^{-2/3}$. There are also 
$\apm$ corrections which arise at string tree level. These correct the
prepotential which determines the leading low-energy action, and are
suppressed by inverse powers of the string frame volume of $M$,
$V_{str}=R^3V_M$. Corrections arise at three-loops, four-loops and 
nonperturbatively in the string sigma model \cand. The leading three-loop
correction to the entropy is supressed by a factor $D^{1/3}/q_0$.
Consistency of the present analysis requires that this be smaller  
than the string loop suppression, or $q_0 \gg D$. This is the same as 
the condition $q_0\gg c_L$ needed in the microscopic analysis.

It would be of interest to go beyond the present analysis and understand 
the leading $\apm$ correction, which amounts to an effective 
shift of $c_2\cdot P/24$ to $q_0$. This was interpreted in the 
string theory picture in \jmcy\ as arising from the anomalous
zerobrane charge of a fourbrane in curved space \refs{\vff, \hmg}.  
In the M-theory 
picture it has the right form to arise from the ground state 
energy of the left-moving chiral bosons. Because there are 
$c_L=6D+c_2\cdot P$ such bosons, 
this is potentially given by  $(6D+c_2\cdot P)/24$ rather
than $c_2\cdot P/24$. In order to compute this shift one must know 
the boundary conditions.  For reasons discussed in \wfv, these scalars
can in general change by additive constants
in going around $S^1$, but from \wfv\ it follows that
this effect alone can not shift $q_0$ appropriately. Indeed 
independent analyses using anomaly inflow 
\refs{\cl, \mr} have recently found an extra shift proportional to 
$D$.
It would be interesting to understand this in the context of the 
semiclassical black hole
entropy formula.

\centerline{\bf Acknowledgements}

We would like to thank D. Morrison and C. Vafa for useful discussions,
and the Aspen Center for Physics, where this 
work was initiated, for hospitality.
J. M. would like to  thank T. Gomez for an initial
 collaboration on this 
project.
This work was supported in part by DOE grant DE-FG02-96ER40559.

\appendix{A}{Normalization of the charge quantization conditions}

Let us fix the normalization of the three form
potential $A^{(3)}$
so that its coupling to the membrane is 
\eqn\twobr{
\int_{\Sigma_3} A^{(3)},
}
and normalize the self-dual field strength $h$ so that its 
coupling on the fivebrane worldvolume  to the three form potential is 
\eqn\braneint{
\int_{\Sigma_6} A^{(3)} \wedge h
.}
The stress tensor on the fivebrane will contain a contribution from $h$
\eqn\stress{
T_{\mu\nu} = B  h_{\mu \rho \sigma} h_\nu^{~\rho \sigma},
}
where $B$ is a constant to be determined.
(Notice that $h_{\mu\rho \sigma} h^{\mu \rho \sigma} =0$ due
to the self-duality condition, so $T_\mu^\mu=0$ as required for
conformal
invariance). 
In order to determine $B$,  we  consider the case of the torus 
and use the known BPS mass formula.
Consider a bound state of a
membrane and a fivebrane represented as an $h$ flux in the fivebrane.
The membrane is along 12 and the fivebrane along 12345. 
The mass of this bound state is
\eqn\massbound{
M = \sqrt{ M_5^2 + M_2^2 } \sim M_5 + {1 \over 2} { R_1 R_2
\over (R_3R_4R_5) },
}
where we used the relation $ T_2^2/T_5 = 2\pi $ obtained in \schwarz .
We write $A^{(3)} = A_\mu dx^\mu \beta_{12}dx^1dx^2 $ where 
$\int_{T_{12} } \beta_{12} = 1 $. 
In this case the coupling \braneint\
becomes 
\eqn\branecou{
\int A_{0}\,\, dt \,\,h_{345} (2\pi)^3 R_3 R_4 R_5 
,}
where the relation between forms and components is 
$ h = { 1\over 6 } h_{\mu\nu\rho} dx^\mu d x^\nu d
x^\rho $.
If $h$ induces one unit of charge we should have
\eqn\valh{
h_{345} = { 1 \over (2 \pi)^3 R_3 R_4 R_5 }.
}
This can be seen by comparing \branecou\ to \twobr .
Inserting \valh\ in \stress\ and comparing it to \massbound\ we
find that
\eqn\valb{
B = { \pi  \over 2 }.
}

Now we return to our case with a generic  Calabi Yau and write 
\eqn\defa{
A^{(3)} = A^A_\mu dx^\mu \alpha_A,
}
where the integral of $\alpha_A$ over the corresponding cycle is
normalized to one. 
Using the formula \hed\ for $h$ and the coupling \asr\ with \defa\
we conclude that the membrane charge is \mch , using 
$k_{\pm}^A \equiv \partial_{\pm} \phi^A $.
Evaluating the momentum is straightforward from  \stress . 
Defining $x^\pm = t \pm x^{11} $ we have
$T_{01} = T_{++} - T_{--}$. 
Using this relation and \hed\ we finally obtain \zmm .

\listrefs

\bye